\documentclass{jfm}
\usepackage{graphicx}
\usepackage{epstopdf, epsfig}
\usepackage[colorlinks = true,
            linkcolor = blue,
            urlcolor  = blue,
            citecolor = blue,
            anchorcolor = blue]{hyperref}

\usepackage[dvipsnames]{xcolor}
\usepackage{amsmath}
\usepackage{bm}
\usepackage[normalem]{ ulem }
\usepackage{soul}
\usepackage{textcomp} 
\usepackage{float}
\usepackage{color}
\usepackage{gensymb}
\usepackage{tikz}
\usepackage{caption}
\usepackage{array,multirow,makecell}
\usepackage{enumitem}
\usepackage{wrapfig}
\usepackage{titlesec}
\usepackage{lmodern}
\usepackage{cases}
\usepackage{siunitx}
\usepackage{csquotes}
\usepackage{empheq}

\newcommand{\pd}{\right)}
\newcommand{\pg}{\left(}
\newcommand{\md}{\right>}
\newcommand{\mg}{\left<}
\newcommand{\cd}{\right]}
\newcommand{\cg}{\left[}

\newcommand{\pa}{\partial}
\newcommand{\nab}{\mathbf{\nabla}}

\newcommand{\son}{c_0}
\newcommand{\dens}{\rho_0}
\newcommand{\pres}{p_0}
\newcommand{\visc}{\mu}
 
\newcommand{\refr}{\mathcal{R}}
\newcommand{\vit}{U}
\newcommand{\vitv}{\mathbf{U}}
\newcommand{\mach}{M}
\newcommand{\surf}{\mathcal{S}}
\newcommand{\norm}{\mathbf{n}}
\newcommand{\sinf}{\mathcal{S_\infty}}
\newcommand{\ninf}{\mathbf{n_\infty}}
\newcommand{\rinf}{r_{\infty}}
\newcommand{\vol}{\mathcal{V}}

\newcommand{\vel}{\mathbf{v}}
\newcommand{\vun}{\mathbf{v_1}}
\newcommand{\vdeux}{\mathbf{v_2}}
\newcommand{\press}{p}
\newcommand{\pun}{p_1}
\newcommand{\pdeux}{p_2}
\newcommand{\dun}{\rho_1}
\newcommand{\ddeux}{\rho_2}
\newcommand{\force}{\mathbf{F}_{\text{rad}}}
\newcommand{\sig}{\Bar{\Bar{\sigma}}}
\newcommand{\ide}{\Bar{\Bar{I}}}

\newcommand{\rb}{r_b}

\shorttitle{Self radiation force}
\shortauthor{A. Roux, J.P. Martishang, M. Baudoin}

\title{Self radiation force on a moving monopolar source}

\author{A. Roux \aff{1}, J.-P. Martischang\aff{1}, M. Baudoin\aff{1} \aff{2} \corresp{\email{michael.baudoin@univ-lille.fr}} }

\affiliation{\aff{1}Univ. Lille, CNRS, Centrale  Lille, Univ. Polytechnique Hauts-de-France, UMR 8520, IEMN, F59000 Lille, 
France \aff{2} Institut Universitaire de France, 1 rue Descartes, 75005 Paris}%

\begin{document}

\maketitle

\begin{abstract}
The radiation force exerted on an object by an acoustic wave is a widely studied phenomenon since the early work of Rayleigh, Langevin and Brillouin and has led in the last decade to tremendous developments for acoustic micromanipulation. Despite extensive work on this phenomenon, the expressions of the acoustic radiation force applied on a particle have so far been derived only for a steady particle, hence neglecting the effect of its displacement on the radiated wave. In this work we study the acoustic radiation force exerted on a monopolar source translating at a constant velocity small compared to the sound speed. We demonstrate that the asymmetry of the emitted field resulting from Doppler effect induces a radiation force on the source opposite to its motion. 
\end{abstract}


\section{Introduction}

Since the seminal work by \cite{pm_rayleigh_1902,pm_rayleigh_1905} and Langevin (work reported later by \cite{ra_biquard_1932a,ra_biquard_1932b}), much effort has been devoted to the derivation of theoretical expressions of the acoustic radiation force exerted by an acoustic wave on a particle. \cite{ap_brillouin_1925,jpr_brillouin_1925} was the first to recognize the tensorial nature of the acoustic radiation force, which is not necessarily orthogonal to the insonified interface. Later on, \cite{king1934acoustic} derived an expression of the axial acoustic radiation force exerted on a rigid spherical particle by a plane wave. This expression was extended to the case of a compressible fluid sphere and of an elastic particle by \cite{a_yosika_1955} and \cite{jasa_hasegawa_1969} respectively. The case of an incident focused wave was treated by \cite{jasa_embleton_1954} for a rigid particle and  \cite{jasa_chen_1996} for an elastic particle. Later on, the more general case of the axial force exerted by a Bessel beam was addressed by \cite{marston2006axial,jasa_marston_2009}.

In parallel, a general expression of the acoustic radiation force exerted by an arbitrary wavefield on a spherical particle in the Long Wavelength Regime (LWR i.e. when $ka \ll 1$ with $k$ the wavenumber and $a$ the particle radius) was obtained by  \cite{spd_gorkov_1962}. It was shown up to 3rd order (in $ka$) that the radiation force is proportional to  the gradient of an acoustic potential, which is proportional to the difference between the time-averaged potential and kinetic acoustic energy weighed respectively by the monopole and dipole scattering coefficients. This expression was extended to the 6th order by \cite{sapozhnikov2013radiation}, which is necessary when the average potential and kinetic energy are uniform in space (e.g. for plane propagating waves). This expression was also extended by \cite{jasa_doinikov_1997a,jasa_doinikov_1997b,jasa_doinikov_1997c} to consider the effect of the viscous and thermal boundary layers in some asymptotic limits (of the boundary layer size compared to the particle size) and in the general case by \cite{pre_settnes_2012} and \cite{pre_karlsen_2015}.  The case of nonspherical particles such as disks and spheroids was treated by \cite{jasa_keller_1957} and \cite{jasa_silva_2018} respectively, while the case of transient acoustic fields was recently addressed by \cite{prap_wang_2021}. We can also note that some expression for the secondary radiation force (inter-particle force) have been derived  by \cite{pre_silva_2014}. The specific case of the radiation force exerted on a vibrating bubble known as the primary Bjerkness force was treated separately by \cite{cup_bjerknes_1906}, \cite{jasa_blake_1949}, \cite{jasa_eller_1968} and \cite{jasa_crum_1975}. Indeed, bubbles have some specificity: owing to their strong compressibility compared to the surrounding liquid, their monopolar resonance appears in the LWR. Hence bubbles can be attracted to the nodes or anti-nodes of a standing wave depending whether they are forced below or above their monopolar resonance frequency (see \cite{jasa_eller_1968}).

Recently, there have been some renewed interest in the calculation of the acoustic radiation force with the development of selective acoustical tweezers (see \cite{arfm_baudoin_2020} for a review of the subject). Acoustical tweezers rely on the acoustic radiation force to move objects. To reach selectivity, i.e. the ability to manipulate a single object independently of other neighboring objects, it is necessary to localize the acoustic energy close to the target particle to only affect it (see e.g. \cite{baresch2016observation}, \cite{sa_baudoin_2019},\cite{baudoin2020naturecell}). Hence, such selectivity cannot be reached in the LWR. In addition, to calculate the restoring force (i.e. the force which brings back the particle toward the trap center), it is necessary to compute the radiation force when the particle is out-centered from the trap position. Yet, all the aforementioned expressions were either limited to the calculation of the radiation force for an axisymmetric configuration or to the LWR. To cope with this issue, general expressions of the radiation force exerted by an arbitrary incident field on a spherical particle, without restriction on the particle size compared to the wavelength, were obtained by \cite{sapozhnikov2013radiation} with an angular spectrum based method
and by \cite{jasa_silva_2011}, and \cite{baresch2013three} with a multipole expansion method. The equivalence between the formula obtained with the different approaches was demonstrated by \cite{jasa_gong_2021}. Note also that some general expressions have been proposed to compute the acoustic radiation torque exerted by an arbitrary acoustic field on a particle of arbitrary size by \cite{epl_silva_2012} and \cite{jasa_gong_2020}.

Yet, in all the theoretical developments mentioned so far, the calculation of the radiation force is made for a steady particle, hence neglecting the effect of its motion. Some account of the interaction between the oscillatory and translational motion of bubbles can be found in the literature. Following \cite{jfm_saffman_1967}, \cite{jfm_benjamin_1990} showed that even in an inviscid fluid, self-propulsion of a bubble can be achieved by nonlinear interactions between adjacent surface deformation modes. In their work however, the surface modes deformation are supposed to be known \textit{a priori}. This work was extended later on by \cite{jfm_mei_1991} to account for the parametric excitation of surface modes by the isotropic volume mode and \cite{pf_feng_1995} who considered the direct coupling between translational motion, volume and shape modes.  Finally, \cite{jfm_doinikov_2004} obtained an expression of this coupling whatever the shape modes, their natural frequency, and the type of excitation (parametric forcing by the volume mode or direct excitation through externally induced pressure gradients at the surface of the bubble). In parallel it was also shown by \cite{pof_watanabe_1993} and \cite{pof_doinikov_2002} that even if the bubble oscillations remain spherical, some complex coupling between volumetric oscillations and translational motion leading to erratic motion of the bubble can still occur when the bubble is excited by acoustic standing waves of high intensity. Finally, we can mention the work of \cite{pof_magnaudet_1998} who computed how the viscous drag applied on a translating bubble is modified by its oscillation.  

\begin{figure}
	\centering
	\includegraphics[width=0.8\textwidth]{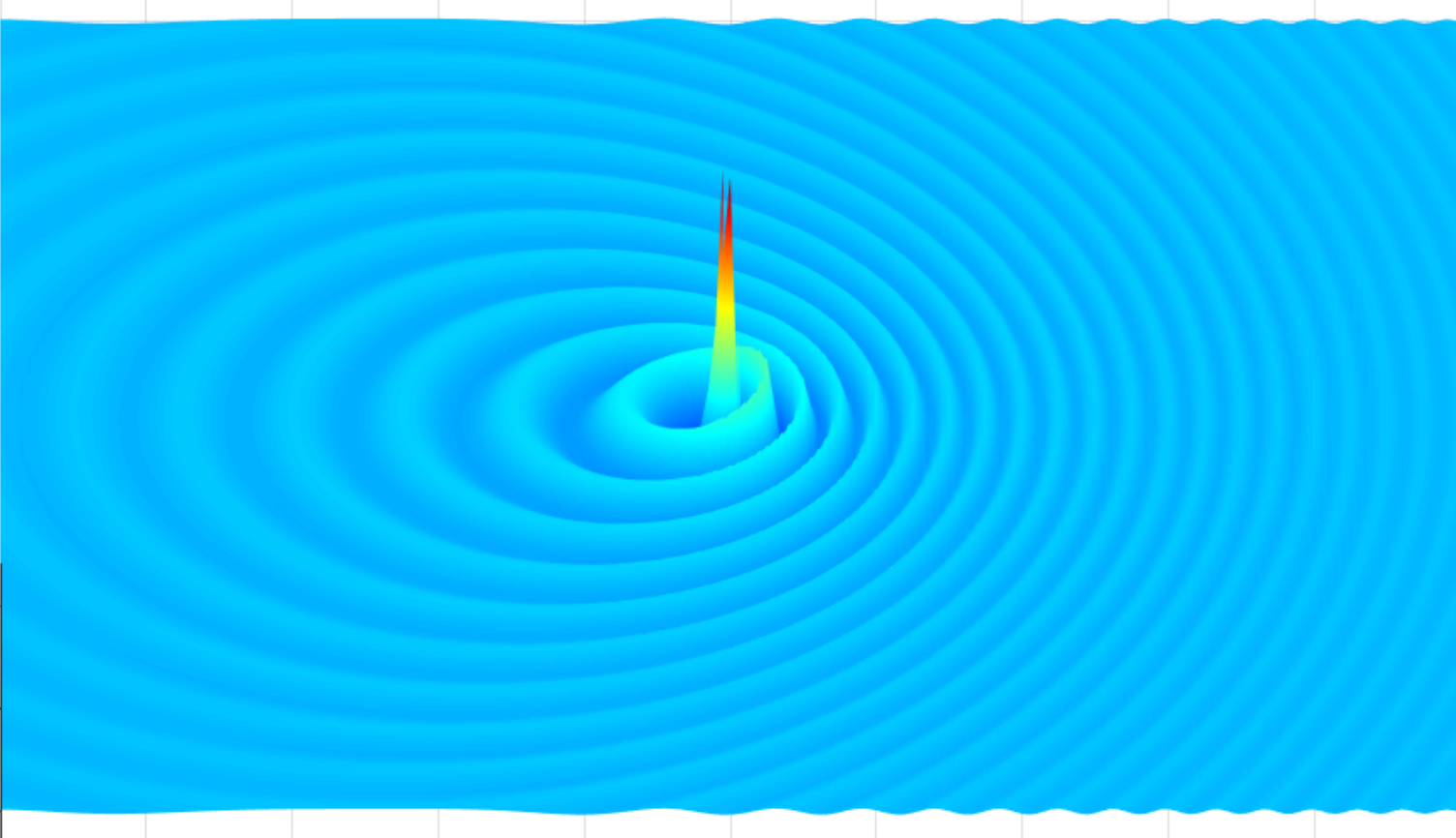}
	\caption{Sketch illustrating the asymmetry of the acoustic field synthesized by a translating monopolar source. The normalized field is calculated with eq. \eqref{eq:ac_pot} and for the sake of illustration, the asymmetry is magnified by choosing a Mach number M = 0.5.}
	\label{sketch}
\end{figure}

But none of these works considered the effect of the asymmetry of the acoustic wave radiated by a translating source on the acoustic radiation force. In this paper, we consider a monopolar source translating in a quiescent inviscid fluid at a constant velocity $U$ along a fixed axis and  demonstrate that the asymmetry of the acoustic field due to Doppler effect (Figure \ref{sketch}) induces a self-induced radiation force on the source resisting its motion. This result is obtained by inserting the well known solution of the wavefield radiated by a moving monopolar source into a far-field integral expression of the radiation force exerted on a moving source, and finally computing this integral within the approximation of slow translating speed compared to the sound speed.

\section{Wavefield radiated by a translating monopolar source}
The first step to compute the self-induced radiation force  exerted on a moving monopolar source is to compute the wavefield radiated by this source in a fixed reference frame. This classic calculation can be found in the acoustics textbook of \cite{mgbc_morse_1968}. In this first section, we recall the main steps of the derivation. Here we suppose the fluid to be inviscid.

\subsection{Wave equation for a translating monopolar source}

In acoustics, a monopolar source can be seen as a source of mass, whose strength is specified by the instantaneous mass flow rate $q(t)$ created by this source. In the following, the source is supposed to be periodic of period T. For a punctual source translating at a velocity $\vitv = \vit \mathbf{x} = M c_o \, \mathbf{x}$ along a fixed axis $\mathbf{x}$, the mass and momentum conservation equations become:
\begin{eqnarray}
& & \frac{\pa\rho}{\pa t}+\nab\cdot\pg\rho\vel\pd= q(t)\delta(x-\mach\son t)\delta(y)\delta(z), \label{eq:mass}\\
& & \frac{\pa\rho\vel}{\pa t} +\nab\pg\rho\vel\otimes\vel\pd=
	\nab\sig \label{eq:mom},
\end{eqnarray}	
with $\refr = (O,(x,y,z),t)$ a Galilean reference frame, $M$ the mach number, $c_o$ the sound speed, $\rho$ the density, $\vel$ the fluid velocity, $\sig$ the stress tensor equal to $-p \ide$ for an inviscid fluid, $\ide$ the identity tensor and $p$ the pressure. If we (i) make the classic asymptotic development of equations \eqref{eq:mass} and \eqref{eq:mom} up to first order:
\begin{subnumcases}{}
\rho = \rho_0 + \epsilon \, \rho_1 \\
p = p_0 + \epsilon \, p_1 \\
\vel = \vel_o + \epsilon \, \vel_1
\end{subnumcases}
with $\epsilon \ll 1$, and obtain the linearized mass and momentum balance:
\begin{eqnarray}
	& & \frac{\pa\dun}{\pa t}+\dens\nab\cdot\vun=q(t)\delta(x-\mach\son t)\delta(y)\delta(z), \label{eq:lin_continuty} \\
	& & \dens\frac{\pa\vun}{\pa t}=-\nab\pun, \label{eq:lin_mom_cons}
\end{eqnarray}
(ii) introduce the sound speed: 
\begin{equation}
\son^2 = \pg\frac{\pa\press}{\pa\rho}\pd_s = \frac{\pun}{\dun}, \label{eq:son}
\end{equation}
where $s$ is the entropy and (iii) combine the time derivative of equation (\ref{eq:lin_continuty}) with the divergence of equation (\ref{eq:lin_mom_cons}), we obtain the wave equation:
\begin{equation}
	\Delta\pun -\frac{1}{\son^2}\frac{\pa^2\pun}{\pa t^2}= 
	-\frac{\partial}{\partial t} \left[ q(t)\delta(x-\mach\son t)\delta(y)\delta(z) \right].
	\label{eq:p}
\end{equation}
Note that to ease the resolution of this problem, it is convenient to introduce the velocity potential $\psi_1$ defined by $\vun=-\frac{1}{\dens}\nab\psi_1$ such that $\pun=\frac{\pa \psi_1}{\pa t}$, which enables to suppress the time derivative in the rhs of equation (\ref{eq:p}): 
\begin{equation}
	\Delta\psi_1 -\frac{1}{\son^2}\frac{\pa^2\psi_1}{\pa t^2}= 
	-q(t)\delta(x-\mach\son t)\delta(y)\delta(z),
	\label{eq:psi}
\end{equation}
and will make the future change of variables easier.
\subsection{Resolution of the wave equation and Lorentz transformation}

The solution of the wave equation (\ref{eq:psi}) is well-known for a fixed monopolar source ($M=0$):
\begin{equation}
	\psi_1(r,t)= \frac{q(t\pm r/\son )}{4\pi r},
\end{equation}
with $r = \sqrt{x^2 + y^2 + z^2}$ the radial distance. To solve the problem for the moving source, the idea is to rewrite equation (\ref{eq:psi}) in a reference frame wherein the source is fixed and the wave equation remains unchanged. This can be achieved by using
the invariance of the wave equation by the Lorentz transformation, which is at the core of special relativity:
\begin{subnumcases}{}
	x'=\gamma\pg x -\mach\son t\pd\\
	y'=y\\
	z'=z\\
	\son t'=\gamma\pg\son t -\mach x\pd,
\end{subnumcases}
where $\gamma$ defined by $\gamma^{-1}=\sqrt{1-\mach ^2}$ is the "Lorentz acoustic boost". With this transformation, the wave equation (\ref{eq:psi}) becomes:
\begin{equation}
	\Delta'\psi_1 -\frac{1}{\son^2}\frac{\pa^2 \psi_1}{\pa t'^2}= 
	-q\pg\gamma(t'+\mach x'/c)\pd\delta(x'/\gamma)\delta(y')\delta(z').
	\label{eq:psip}
\end{equation}
Since the rhs of equation (\ref{eq:psip}) is null when $x'\neq 0$ for all $t'$ and using $\delta(x'/\gamma)=\gamma\delta(x')$, it comes:	
\begin{equation}
	\Delta' \psi_1 -\frac{1}{\son ^2}\frac{\pa^2 \psi_1}{\pa t'^2}= 
	-\gamma q(\gamma t')\delta(x')\delta(y')\delta(z').
\end{equation}
If we now introduce a second set of variables:
\begin{subnumcases}{}
	x''=\gamma x'\\
	y''=\gamma y'\\
	z''=\gamma z'\\
	\son t''=\gamma \son t',
\end{subnumcases}
the wave equation becomes:
\begin{equation}
	\Delta'' \psi_1 -\frac{1}{\son ^2}\frac{\pa^2 \psi_1}{\pa t''^2}= 
	-\gamma^2 q(t'')\delta(x'')\delta(y'')\delta(z''),
\end{equation}	
which now resembles the static monopolar source problem and whose solution is:	
\begin{equation}
	\psi_1(r'',t'')= \gamma^2\frac{q(t''\pm r''/\son )}{4\pi r''}.
\end{equation}
If we now perform the inverse transformations to obtain the potential as a function of $(x,y,z,t)$, we obtain:
\begin{equation}
	\psi_1(r',t')= \gamma^2\frac{q(\gamma(t'\pm r'/\son ))}{4\pi\gamma r'},
\end{equation}
with:
\begin{eqnarray}
	\gamma(t'\pm \frac{r'}{\son}) & =  &
	\frac{\gamma}{\son}\cg\gamma(\son t-\mach x) \pm \sqrt{[\gamma(x-\mach\son t)]^2+y^2+z^2}\cd \\
	 & =  & t -\frac{\mach (x-\mach\son t)\pm\sqrt{(x-\mach\son t)^2+(y^2+z^2)(1-\mach ^2)}}{\son (1-\mach ^2)},
\end{eqnarray}
If we introduce the distance $R_\pm$ between the emission and the observation points:
\begin{equation}
	R_\pm=\frac{\mach(x-\mach \son t)\pm R_1}{1-M^2},
\end{equation}
with 
\begin{equation}
	R_1=\sqrt{(x-\mach \son t)^2+(y^2+z^2)(1-M^2)},
\end{equation}
then the solution becomes:
\begin{equation}\label{eq:ac_pot}
\boxed{
	\psi_1(r,t)=\frac{q(t-R/\son )}{4\pi R_1},
	}
\end{equation}
with $R = R_+$ in the subsonic case ($\mach <1$).

\section{Integral expression of the radiation stress in the far field}

The next step is to derive a far-field integral expression of the radiation stress exerted on a moving source. Indeed in acoustics, a monopolar point source constitutes a far-field approximation of a real source of finite extent and hence the above expressions are only valid in the far field.

\begin{figure}
	\centering
	\begin{tikzpicture}[scale=1]
		\draw[->,color=BurntOrange] (0,0)--++(6,0) node [above] {$x$};
		\draw[->,color=BurntOrange] (0,0)--++(0,3) node [right] {$y$};
		\draw[->,color=BurntOrange] (0,0)--++(-2,-2) node [below right] {$z$};
		\node[color=BurntOrange] at (-1,1.3) []{$\refr$};
		\draw[->,color=TealBlue] (3,0)--++(4,0) node [above] {$x^*$};
		\draw[->,color=TealBlue] (3,0)--++(0,3) node [right] {$y^*$};
		\draw[->,color=TealBlue] (3,0)--++(-2,-2) node [below right] {$z^*$};
		\node[color=TealBlue] at (2,1.3) []{$\refr^*$};
		
		\draw [color=white,ball color=gray,smooth] (3,0) circle (0.5);
		\draw [->,line width=1.2 pt,color=DarkOrchid] (3,0) --++ (1.5,0) node [below left] {$U$};
		\node[] at (3.5,-0.5) []{$\surf$};
		\draw [->,line width=1.2 pt,color=ForestGreen] (3.4,0.3) --++ (0.5,0.5) node [below right] {$\norm$};
		\draw [->,line width=1.2 pt,color=ForestGreen] (5,1.5) --++ (0.5,0.5) node [below right] {$\ninf$};
			
		\draw [-] (3,0) circle (2.5);
		\draw [-] (5.5,0) arc (0:-180:2.5 and 1) ;
		\draw [dashed] (5.5,0) arc (0:180:2.5 and 1) ;
		\draw [dashed] (3,2.5) arc (90:270:1.8 and 2.5) ;
		\draw [-] (3,2.5) arc (90:-90:1.8 and 2.5) ;
		\node[] at (5,-2) []{$\surf_\infty$};
	\end{tikzpicture}
	\caption{$\surf$ represents the source surface, varying over time. The surface $\sinf$ is centered on the source and moves with it at the velocity $\mathbf{\vit }$ in $\refr$. $\refr^*$ is the frame of the source.}
	\label{fig:2} 
\end{figure}
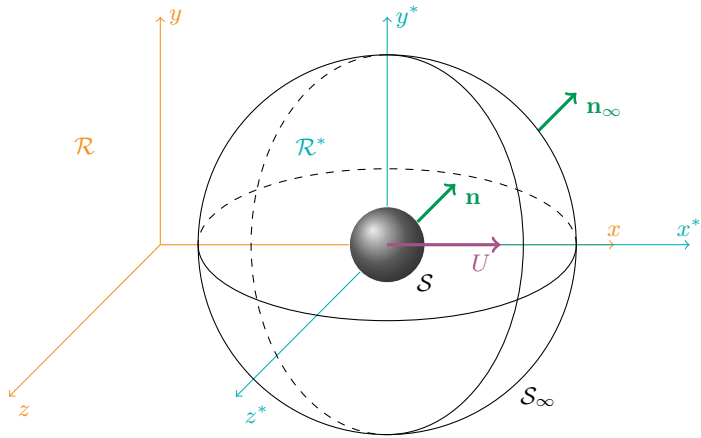

\subsection{Far field expression of the radiation force for a moving source}

The acoustic radiation stress exerted on an object of surface $\surf(t)$ is by definition the time average of the surface integral of the stress exerted by the acoustic wave on its surface:
\begin{equation}
	\mg\force\md=\mg\iint_{\surf(t)}\sig\norm dS\md,
\end{equation}	
where $\mg f \md = \frac{1}{T} \int_t^{t+T} f(t) dt$ is the time average of the function $f$ and T the period of the function $f(t)$. In general, there are two difficulties when computing this integral: (i) the surface of the object is vibrating and hence depends on time ($\surf = \surf(t)$) and (ii) an expression of the wave scattered by the object in the near field must be known. Hence generally this integral is converted into an integral over a closed surface at rest surrounding the object in the far field by using the divergence theorem and Reynolds transport theorem (see e.g. the review by \cite{arfm_baudoin_2020} for details of this process). Here an additional difficulty comes from the fact that the particle, in addition to its vibration, is translating at a constant velocity $U$. To solve this issue, we will transpose our integral of the stress on the surface of the object into an integral over a spherical surface $\sinf$ of radius $\rinf \gg \lambda$, centered on the source, and hence translating at the velocity $\mathbf{\vit}$ in $\refr$ (see Figure \ref{fig:2}), with $\lambda = \son / f $ the wavelength and $f$ the frequency. The volume between $\surf$ and $\sinf$ is named $\vol$. The integral of equation (\ref{eq:mom}) over $\vol$ gives:
\begin{equation}
	\iiint_\vol{\cg\frac{\pa\rho\vel}{\pa t} 
	+\nab\pg\rho\vel\otimes\vel 
	-\sig\pd\cd dV}=
	\mathbf{0}.
\end{equation}
Using the divergence theorem, this volume integral turns into:
\begin{equation}\label{eq:momentum}
	\iiint_\vol{\frac{\pa\rho\vel}{\pa t}dV} -\iint_\surf\pg\rho\vel\otimes\vel-\sig\pd\norm dS +\iint_\sinf\pg\rho\vel\otimes\vel-\sig\pd\ninf dS=
	\mathbf{0},
\end{equation}
with $\norm$ and $\ninf$ the outgoing normal vectors to the surface $\surf$ and $\sinf$ respectively (see Figure \ref{fig:2}). Another equation can be obtained by applying the Reynolds transport theorem to the momentum density $\rho\vel$:
\begin{equation}\label{eq:reynolds}
	\frac{d}{dt}\iiint_\vol{\rho\vel dV}=
	\iiint_\vol{\frac{\pa \rho\vel}{\pa t}dV}
	-\iint_\surf(\vel\cdot\norm)\rho\vel dS
	+\iint_\sinf(\mathbf{\vit }\cdot\ninf)\rho\vel dS,
\end{equation}
since the surface $\surf$ follows the object surface displacement (equal to the fluid displacement at the interface due to the continuity condition) and the surface $\sinf$ is translating at a constant velocity $\vel$. If we study the steady regime, then:
\begin{equation}
	\mg\frac{d}{dt}\iiint_\vol{\rho\vel dV}\md=\mathbf{0}.
	\label{eq:ta}
\end{equation}
Finally, if we substract the time average of equation \eqref{eq:reynolds} to the time average of equation \eqref{eq:momentum} and take into account (\ref{eq:ta}), we obtain:
\begin{equation}\label{eq:force_general}
	\mg\force\md= 
	\mg\iint_\sinf\pg\sig-\rho\vel\otimes\vel\pd\ninf dS\md
	+\mg\iint_\sinf(\mathbf{\vit }\cdot\ninf)\rho\vel dS\md.
\end{equation}

\subsection{Expression as a function of the first order acoustic field}

In order to compute the previous integrals and since the time average of first order terms are equal to $\mathbf{0}$ for a periodic signal, we need to express the terms appearing in equation (\ref{eq:force_general}) up to second order. Here we suppose again that the fluid is inviscid and hence $\sig = - p \ide$. If we push the asymptotic development up to second order:
\begin{subnumcases}{}
		\rho=\dens+\epsilon \dun+ \epsilon^2 \ddeux \\
	\press=\pres+\epsilon \pun+  \epsilon^2 \pdeux\\
    \vel= \epsilon \vun+ \epsilon^2 \vdeux,
\end{subnumcases}
the momentum balance \eqref{eq:mom} at second order in the volume $\vol$ becomes:
\begin{equation}
	\dens\frac{\pa\vdeux}{\pa t}
	+\dun\frac{\pa\vun}{\pa t} 
	+\dens \pg \vun \nab \vun \pd =
	-\nab \pdeux.
\end{equation}
If we take the time-average of this equation, we obtain:
\begin{equation}
	\mg-\nab \pdeux\md=\mg\dun\frac{\pa \vun}{\pa t}+\dens\pg \vun \nab \vun \pd \md,
\label{eq:p2mo}
\end{equation}
since $\mg\frac{\pa \vdeux}{\pa t}\md=\mathbf{0}$ in the steady regime.
From the first order momentum conservation \eqref{eq:lin_mom_cons}, we have $\frac{\pa \vun}{\pa t} = - \frac{1}{\dens} \nabla p_1$ and from the state equation (\ref{eq:son}) we have $\rho_1 = p_1 / c_0^2$ leading to:
\begin{equation}
\dun\frac{\pa \vun}{\pa t} = -\frac{1}{2\dens\son ^2}\nab\pun^2.
\label{eq:pot}
\end{equation}
And since the acoustic field is by definition irrotationnal:
\begin{equation}
\dens\pg \vun \nab \vun \pd = \frac{\dens}{2}\nab\vun^2.
\label{eq:cin}
\end{equation}
Finally, by replacing \eqref{eq:pot} and \eqref{eq:cin} in \eqref{eq:p2mo}, we obtain:
\begin{equation}
	\mg-\nab \pdeux\md=
	\mg-\frac{1}{2\dens\son ^2}\nab\pun^2
	+\frac{\dens}{2}\nab\vun^2\md.
\end{equation}
From this equation, we obtain the second order averaged stress tensor:
\begin{equation}
	\mg \sig_2 \md= -\mg \pdeux \ide \md =
	\mg \pg\dens\frac{v_1^2}{2}
	-\frac{1}{\dens\son^2}\frac{\pun^2}{2}\pd\ide \md. \label{eq:stress2}
\end{equation}
Then at second order, we have:
\begin{eqnarray}
& & \mg \rho\vel\otimes\vel \md = \mg \dens \vun \otimes\vun \md \label{eq:conv2} \\
& & \mg \rho \vel \md = \mg \dun \vun \md \label{eq:m2},
\end{eqnarray}
so that if we replace equations \eqref{eq:stress2}, \eqref{eq:conv2} and \eqref{eq:m2} in \eqref{eq:force_general}, we obtain the final expression of the radiation force as a function of the first order acoustic fields:
\begin{equation}\label{eq:force}
\boxed{
	\mg\force\md=
	\mg\iint_{\sinf}{\cg\pg\dens\frac{v_1^2}{2}-\frac{1}{\dens\son^2}\frac{\pun^2}{2}\pd \ide  -\dens\vun\otimes\vun\cd}\mathbf{n} dS\md
	+\mg\iint_\sinf(\mathbf{\vit }\cdot\mathbf{n_\infty})\dun\vun dS\md.
}
\end{equation}

\section{Expression of the radiation force for a slowly translating source}

The final step is to compute this integral from the first order acoustic field.

\subsection{Pressure, density and velocity fields at first order}

The first step to compute equation \eqref{eq:force} is to determine the pressure, density and velocity fields at first order from the expression of the velocity potential determined previously (equation \eqref{eq:ac_pot}). First,

\begin{equation}
	\pun=\frac{\pa \psi_1}{\pa t}=\frac{(1-\frac{1}{\son}\frac{\pa R}{\pa t})q'(t-R/\son )}{4\pi R_1}
	-\frac{q(t-R/\son )\frac{\pa R_1}{\pa t}}{4\pi R_1^2}.
\end{equation}
Moreover, the coordinate of the velocity $v_{1i}$ along the direction $x_i = (x,y,z)$ is:
\begin{equation}
	v_{1x_i}=-\frac{1}{\dens}\frac{\pa \psi_1}{\pa x_i}=
	-\frac{1}{\dens}\pg\frac{-\frac1\son \frac{\pa R}{\pa x_i}q'(t-R/\son )}{4\pi R_1}
	-\frac{q(t-R/\son )\frac{\pa R_1}{\pa x_i}}{4\pi R_1^2}\pd.
\end{equation}
Of course, we have: $\dun = \pun / \son^2$. Here $R$ and $R_1$ are time and space dependent functions. To simplify the calculation of the derivatives and the force integral, we perform the change of variable corresponding to the Galilean transformation from $\refr$ to $\refr^*$:
\begin{subnumcases}{\label{eq:gal}}
 x^*=x-\mach \son t\\
	y^*=y\\
	z^*=z\\
	t^*=t
\end{subnumcases}
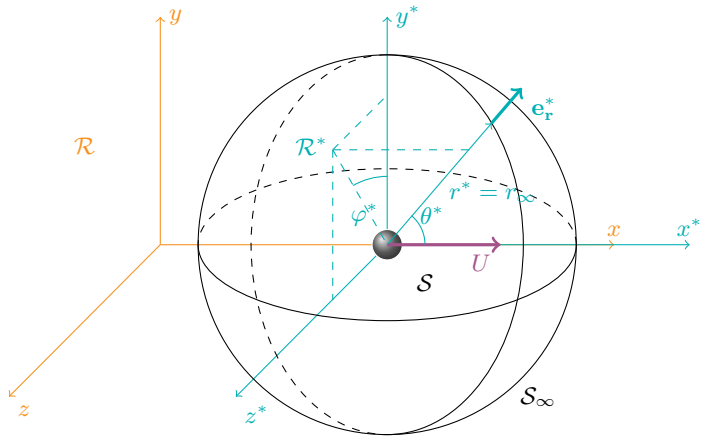
\begin{figure}
	\centering
	\begin{tikzpicture}[scale=1]
		\draw[->,color=BurntOrange] (0,0)--++(6,0) node [above] {$x$};
		\draw[->,color=BurntOrange] (0,0)--++(0,3) node [right] {$y$};
		\draw[->,color=BurntOrange] (0,0)--++(-2,-2) node [below right] {$z$};
		\node[color=BurntOrange] at (-1,1.3) []{$\refr$};
		\draw[->,color=TealBlue] (3,0)--++(4,0) node [above] {$x^*$};
		\draw[->,color=TealBlue] (3,0)--++(0,3) node [right] {$y^*$};
		\draw[->,color=TealBlue] (3,0)--++(-2,-2) node [below right] {$z^*$};
		\node[color=TealBlue] at (2,1.3) []{$\refr^*$};
		
		\draw [color=white,ball color=gray,smooth] (3,0) circle (0.2);
		\draw [->,line width=1.2 pt,color=DarkOrchid] (3,0) --++ (1.5,0) node [below left] {$U$};
		\node[] at (3.5,-0.5) []{$\surf$};
		
		\draw [-] (3,0) circle (2.5);
		\draw [-] (5.5,0) arc (0:-180:2.5 and 1) ;
		\draw [dashed] (5.5,0) arc (0:180:2.5 and 1) ;
		\draw [dashed] (3,2.5) arc (90:270:1.8 and 2.5) ;
		\draw [-] (3,2.5) arc (90:-90:1.8 and 2.5) ;
		\node[] at (5,-2) []{$\surf_\infty$};
		
		\draw [->,color=TealBlue] (3,0) --++ (1.38,1.6);
		\draw [->,line width=1.2 pt,color=TealBlue] (4.38,1.6) --++ (0.4,0.464) node [below right] {$\mathbf{e_{r}^*}$};
		\node[color=TealBlue] at (4.4,0.7) []{$r^*=\rinf$};
		\draw [dashed,color=TealBlue] (4.08,1.25) --++ (-1.8,0);
		\draw [dashed,color=TealBlue] (2.28,1.25) --++ (0.72,-1.25);
		\draw [dashed,color=TealBlue] (2.28,1.25) --++ (0.7,0.7);
		\draw [dashed,color=TealBlue] (2.28,1.25) --++ (0,-1.98);
		\draw [color=TealBlue] (3.5,0) arc (0:50:0.5) node [right] {$\theta^*$};
		\draw [color=TealBlue] (3,0.9) arc (90:120:0.9); 
		\node[color=TealBlue] at (2.7,0.4) []{$\varphi^*$};
	\end{tikzpicture}
	\caption{We make the change of variables corresponding to the Galilean transformation from $\refr$ to $\refr^*$ and then use the local spherical coordinates $(r^*,\theta^*,\varphi^*)$.}
	\label{fig:3} 
\end{figure}
$R_1$ and $R$ become:
\begin{equation}
	R_1=\sqrt{{x^*}^2+({y^*}^2+{z^*}^2)(1-\mach ^2)} \quad\text{and}\quad
	R=\frac{\mach x^*+R_1}{1-M^2}.
\end{equation}
The Jacobian matrix $J^*$ of this transformation is:
\begin{equation}
	J^*=
	\begin{pmatrix}
		1 & 0 & 0 & -M \son \\
		0 & 1 & 0 & 0\\
		0 & 0 & 1 & 0\\
		0 & 0 & 0 & 1
	\end{pmatrix}
	\quad\text{with}\quad
	\det J^*=1.
\end{equation}
We use the spherical coordinates $(r^*, \theta^*, \varphi^*)$ to compute the integral (see Figure \ref{fig:3} for the notations):
\begin{subnumcases}{}
	x^*=r^*\cos\theta^*\\
	y^*=r^*\sin\theta^*\cos\varphi^*\\
	z^*=r^*\sin\theta^*\sin\varphi^*.
\end{subnumcases}
The key point in the following derivation of the self radiation force is that we consider $\mach\ll1$, which means that the source translates at low velocity compared to the sound speed. This enables us to get simpler expressions of the pressure and velocity fields. First, we derive the expressions of $R_1$ and $R$ at the first order in $M$:
\begin{align}
	R_1 & =\sqrt{{r^*}^2-\mach ^2({y^*}^2+{z^*}^2)}=
	\sqrt{{r^*}^2-\mach ^2{r^*}^2\sin^2\theta^*} \nonumber \\
	& \simeq r^*\pg1-\frac{\mach ^2}{2}\sin^2\theta^*\pd\simeq r^*+\mathcal{O}(M),
\end{align}
and
\begin{equation}
	R\simeq\frac{\mach x^*+r^*\pg1-\frac{\mach^2}{2}\sin^2\theta^*\pd}{1-\mach^2}\simeq r^*+Mx^*.
\end{equation}
In order to get the velocity and the pressure fields at the first order in $\mach$, we compute at first the different time derivatives of $R_1$ and $R$:
\begin{equation}
	\frac{\pa R}{\pa t}=
	\frac{\pa x^*}{\pa t} \frac{\pa R}{\pa x^*}+\frac{\pa t^*}{\pa t}\frac{\pa R}{\pa t^*}=
	-\mach \son \pg\frac{\pa r^*}{\pa x^*}+M\pd,
\end{equation}
with:
\begin{equation}
	\frac{\pa r^*}{\pa x^*}=\frac{x^*}{r^*} = \cos \theta^*.
\end{equation}
Hence we obtain:
\begin{equation}
	\frac{\pa R}{\pa t}= -\mach \son \pg\cos\theta^*+M\pd.
\end{equation}
Then:
\begin{equation}
	\frac{\pa R_1}{\pa t}=
	\frac{\pa x^*}{\pa t} \frac{\pa R_1}{\pa x^*} 
	+\frac{\pa t^*}{\pa t}\frac{\pa R_1}{\pa t^*}=	-\mach \son \cos\theta^*.
\end{equation}
Finally, the acoustic pressure is:
\begin{equation}
	\pun=\frac{\cg1+\mach \pg\cos\theta^*+\mach \pd\cd q'(t-R/\son )}{4\pi r^*}
	+\frac{q(t-R/\son )(\mach \son \cos\theta^*)}{4\pi {r^*}^2}.
\end{equation}
We choose a sphere $\sinf$ with a radius $\rinf$ huge compared to all the other lengths of the problem. We then write the acoustic pressure in the far field approximation. At first order in $\mach $ we obtain:
\begin{equation}
\boxed{
	\pun=\frac{1+\mach \cos\theta^*}{4\pi r^*}q'(t-R/\son ).
	}
\end{equation}
In order to derive the velocity field we compute the different space derivatives of $R_1$ and $R$:
\begin{subnumcases}{}
	\frac{\pa R}{\pa x}=\frac{\pa x^*}{\pa x} \frac{\pa R}{\pa x^*}=\cos\theta^*+\mach \\
	\frac{\pa R}{\pa y}=\frac{\pa y^*}{\pa y} \frac{\pa R}{\pa y^*}=\sin\theta^*\cos\varphi^*\\
	\frac{\pa R}{\pa z}=\frac{\pa z^*}{\pa z} \frac{\pa R}{\pa z^*}=\sin\theta^*\sin\varphi^*,
\end{subnumcases}
and 
\begin{equation}
	\frac{\pa R_1}{\pa xi}= \frac{x_i^*}{r^*}.
\end{equation}
We obtain:
\begin{subnumcases}{}
	v_{1x^*}=\frac{1}{\dens\son}\pg\frac{(\cos\theta^*+\mach )q'(t-R/\son )}{4\pi r^*}+\son\frac{q(t-R/\son )}{4\pi {r^*}^2}\frac{x^*}{r^*}\pd\\
	v_{1y^*}=\frac{1}{\dens\son}\pg\frac{\sin\theta^*\cos\varphi^* q'(t-R/\son )}{4\pi r^*}+\son\frac{q(t-R/\son )}{4\pi {r^*}^2}\frac{y^*}{r^*}\pd\\
	v_{1z^*}=\frac{1}{\dens\son}\pg\frac{\sin\theta^*\sin\varphi^* q'(t-R/\son )}{4\pi r^*}+\son\frac{q(t-R/c)}{4\pi {r^*}^2}\frac{z^*}{r^*}\pd.
\end{subnumcases}
We finally write the velocity field in the far field approximation:
\begin{empheq}[box=\fbox]{align}
	&v_{1x^*}=\frac{1}{\dens \son }\frac{q'(t-R/\son )}{4\pi r^*}(\cos\theta^*+\mach ) \label{vux} \\
	&v_{1y^*}=\frac{1}{\dens \son }\frac{q'(t-R/\son )}{4\pi r^*}\sin\theta^*\cos\varphi^*\\
	&v_{1z^*}=\frac{1}{\dens \son }\frac{q'(t-R/\son )}{4\pi r^*}\sin\theta^*\sin\varphi^*. \label{vuz}
\end{empheq}
\subsection{Calculation of the integrals}
With the change of variables \eqref{eq:gal} and since the Jacobian of this change of variable is equal to 1, the integral expression of the radiation force \eqref{eq:force} becomes:
\begin{equation}\label{eq:forcef}
	\mg\force\md=
	\mg\iint_{\sinf^*}{\cg\pg\dens\frac{v_1^2}{2}-\frac{1}{\dens\son^2}\frac{\pun^2}{2}\pd \ide  -\dens\vun\otimes\vun\cd}\mathbf{n^*} dS^*\md
	+\mg\iint_{\sinf^*}{\vit }\cdot\mathbf{n_\infty}^*)\dun\vun dS^*\md,
\end{equation}
with $\sinf^*$ the surface defined by $r^*=\rinf^*$ a constant radius, the  infinitesimal element of surface $dS^*={\rinf^*}^2\sin\theta^* d\theta^* d\varphi^*$, $\theta^* \in \left[0,\pi \right]$ and  $\varphi^*\in \left[0, 2 \pi \right]$. Finally, $\mathbf{n^*} = \ninf^* =\mathbf{e_{r}^*}$, with $(\mathbf{e_{r}^*},\mathbf{e_{\theta}^*},\mathbf{e_{\varphi}^*})$ the spherical coordinates unit vector of $\refr^*$. Since we have already expressed the pressure, velocity and density fields as a function of the spherical coordinates $(r^*, \, \theta^*, \, \varphi^*)$, we can now compute integral \eqref{eq:forcef}. In the following paragraphs, we compute separately each term of this integral.
\subsubsection{Potential energy term}
\begin{align}
	& \mg\iint_{\sinf^*}{\frac{1}{\dens \son ^2}\frac{\pun^2}{2}}\mathbf{n^*}dS^*\md \nonumber \\
	& =
	\mg\frac{1}{2\dens\son^2}\int_{\varphi^*=0}^{2\pi}\int_{\theta^*=0}^{\pi}{\left[ \pg\frac{1+\mach \cos\theta^*}{4\pi {\rinf^*}}\pd q'(t-R/\son ) \right]^2}\mathbf{e_{r}^*}{\rinf^*}^2\sin\theta^* d\theta^* d\varphi^*\md
\end{align}
The integration over $\varphi^*$ along the $y^*$ and $z^*$ axis is null. Hence only the term along $x^*$ remains:
\begin{equation}
	\mg\iint_{\sinf^*}{\frac{1}{\dens \son ^2}\frac{\pun^2}{2} }\mathbf{n^*}dS^*\md=
	\mg\frac{\pi}{\dens\son^2}\int_{0}^{\pi}{\left[ \pg\frac{1+\mach \cos\theta^*}{4\pi {\rinf^*}}\pd q'(t-R/\son ) \right]^2}\cos\theta^*\sin\theta^*d\theta^* \mathbf{x} \md .
\end{equation}
If we swap the time and space integrals and since the function $q(t)$ is periodic, we obtain:
\begin{equation}
	\mg\iint_{\sinf^*}{\frac{1}{\dens\son^2}\frac{\pun^2}{2}}\mathbf{n^*}dS^*\md=
	\frac{\mg q'(t)^2\md}{16\pi\dens\son^2}\int_{0}^{\pi}{\pg1
	+\mach \cos\theta^*\pd^2}\cos\theta^*\sin\theta^* d\theta^* \mathbf{x}.
\end{equation}
Since the last integral is equal to $4\mach/3$, we obtain:
\begin{equation}
\boxed{
	\mg\iint_{\sinf^*}{\frac{1}{\dens\son^2}\frac{\pun^2}{2}}\mathbf{n^*}dS^*\md=
	\frac{\mg q'^2\md \mach}{12\pi\dens\son^2} \mathbf{x}.
			\label{fin:pot}
	}
\end{equation}
\subsubsection{Kinetic energy term}
\begin{align}
	& \mg\iint_{\sinf^*}{\dens\frac{v_1^2}{2}}\mathbf{n^*}dS^*\md \nonumber \\
	& = \mg\frac{1}{2\dens\son^2}\int_{\varphi^*=0}^{2\pi}\int_{\theta^*=0}^{\pi}{\pg\frac{q'(t-R/\son)}{4\pi{\rinf^*}}\pd^2\pg(\cos\theta^*+\mach)^2+\sin^2\theta^*\pd}\mathbf{e_{r}^*}{\rinf^*}^2\sin\theta^*d\theta^*d\varphi^*\md
\end{align}
Using the same arguments than previously we have:
\begin{equation}
	\mg\iint_{\sinf^*}{\dens\frac{v_1^2}{2}}\mathbf{n^*}dS^*\md=
	\frac{\mg q'^2\md}{16\pi\dens \son ^2}\int_0^\pi{\pg 1 + 2 M\cos\theta^* + M^2\pd}\cos\theta^*\sin\theta^*d\theta^*\mathbf{x} 
\end{equation}
Since the last integral is equal to $4\mach/3$, we obtain:
\begin{equation}
\boxed{
	\mg\iint_{\sinf^*}{\dens\frac{v_1^2}{2}}\mathbf{n^*}dS^*\md=
	\frac{\mg q'^2\md \mach}{12\pi\dens\son^2}\mathbf{x}
			\label{fin:kin}
	}
\end{equation}

\subsubsection{Convective term}
Due to the symmetry of the problem (invariance by rotation over $\varphi^*$ angle), no force can exist along $\mathbf{y}$ and $\mathbf{z}$ directions. Hence, we only need to compute the following components of $\vun\otimes\vun$:
\begin{equation}
  (v_{1x^*}^2) \mathbf{x} \otimes \mathbf{x} +(v_{1x^*}v_{1y^*}) \mathbf{x} \otimes \mathbf{y} +(v_{1x^*}v_{1z^*}) \mathbf{x} \otimes \mathbf{z}
\end{equation}
Then due to the dependence of these terms over $\varphi^*$ given in equations \eqref{vux} to  \eqref{vuz} and since $\mathbf{e_r^*} = \cos \theta^* \mathbf{x} + \sin \theta^* \cos \varphi^* \mathbf{y} + \sin \theta^* \sin \varphi^* \mathbf{y} $, we obtain:
\begin{align}
	& \mg\iint_{\sinf^*}{\cg\dens\vun\otimes\vun\cd}\mathbf{n^*} dS^*\md \nonumber \\
	& = \frac{\mg q'^2\md}{8 \pi \dens \son^2} \int_0^\pi{ \left[ \pg\cos\theta^*+\mach\pd^2 \cos\theta^* + \pg\cos\theta^*+\mach\pd {\sin\theta^*}^2 \right]\sin\theta^* d\theta^*}\mathbf{x}
\end{align}
The integral term is equal to $8\mach/3$, so that:
\begin{equation}
\boxed{
	\mg\iint_{\sinf^*}{\cg\dens\vun\otimes\vun\cd}\mathbf{n^*} dS^*\md=
	\frac{\mg q'^2 \md \mach }{3\pi\dens \son ^2}  \mathbf{x}
		\label{fin:conv}
	}
\end{equation}

\subsubsection{Source translation term}
The term due the translation of the sphere $\sinf$ is:
\begin{equation}
	 \mg\iint_{\sinf^*}(\mathbf{\vit }\cdot\mathbf{n_\infty^*})\dun\vun dS^*\md \nonumber  = 	\frac{\mach\mg q'^2\md}{8\pi\dens \son ^2}\int_0^\pi(1+\mach \cos\theta^*)(\cos\theta^*+\mach )\cos\theta^*\sin\theta^* d\theta^*\mathbf{x}
\end{equation}
Since at leading order, the last integral is equal to $2/3$, we finally obtain:
\begin{equation}
\boxed{
	\mg\iint_{\sinf^*}(\mathbf{\vit }\cdot\mathbf{n_\infty})\dun\vun dS^*\md=
	\frac{\mg q'^2\md \mach}{12\pi\dens c^2}   \mathbf{x}
	\label{fin:trans}
	}
\end{equation}
\section{Final expression of the self radiation force and discussion}
\subsection{Final expression of the self-radiation force}
If we now replace \eqref{fin:pot}, \eqref{fin:kin}, \eqref{fin:conv} and \eqref{fin:trans} in \eqref{eq:forcef}, we obtain the final expression of the self acoustic radiation force exerted on a monopolar source:
\begin{equation}
\boxed{
	\mg\force\md=-\frac{\mg q'^2\md \mach}{4\pi\dens \son ^2} \mathbf{x}.
	}
	\label{eq:final}
\end{equation}
There are many interesting things to notice in the above calculations and final expression. First, we see that the potential and kinetic energy terms cancel at leading order in M, so that the self radiation force is solely due to the convective and translation terms. Second, the radiation force is proportional to the radiated intensity, and inversely proportional to the sound speed square, which is classical in radiation force calculations. In addition, here the self-radiation force is proportional to the hydrodynamic Mach number $M$, which is expected since the force results from the asymmetry of the radiated field due to the translation of the source. Finally and most importantly $ \mg\force\md\cdot\mathbf{\vit }$ is always negative, which means that this force always slows down the movement of the bubble.


\subsection{An example of a monopolar oscillator: a vibrating bubble}

In the following section, we estimate this force for a translating and oscillating bubble in a liquid, which constitutes one example of acoustic monopolar source. Indeed, bubbles are exceptional resonators, which exhibit strong monopolar resonances in the long wavelength regime. Let's consider a spherical bubble of mean radius $\rb$ in a liquid of density $\dens$ and sound speed $\son$ vibrating periodically at its resonance frequency, called the Minnaert frequency:
\begin{equation}
	\omega_M=\frac{1}{\rb}\sqrt{\frac{3\gamma\pres}{\dens}},
\end{equation}
with $\gamma$ the heat capacity ratio of the gas in the bubble and $\pres$ the pressure of the surrounding fluid. At resonance, $\lambda/r_b=\frac{2\pi \son}{\omega_M r_b}=2\pi \son \sqrt{\frac{\dens}{3\gamma \pres}} \gg 1$ (of the order of $5 \times 10^2$ for an air bubble in water). Since the bubble is very small compared to the wavelength in this case, it can be considered as a point source a few wavelengths away from the bubble surface. The oscillation of this bubble creates a periodic mass flux  $q(t) = Q \cos(\omega_M t)$, whose magnitude $Q$ is basically equal to the surface of the bubble $4\pi \rb^2$, times the surrounding liquid mass density $\dens$, times the amplitude of the oscillations $\alpha \rb$ (where $\alpha$ designates a dimensionless parameter fixing the magnitude of the bubble oscillation), times the pulsation $\omega_b$:
\begin{equation}
	Q \sim 4\pi \rb^2\dens\alpha \rb \frac{1}{\rb}\sqrt{\frac{3\gamma p_0}{\dens}} = 4 \pi r_b^2 \alpha \sqrt{3 \gamma p_0 \rho_0}.
\end{equation}
Consequently, we have:
\begin{equation}
	\left< q'^2 \right> \sim \frac{1}{2} \omega_M^2 Q^2 = 72 (\pi \gamma p_0 \alpha \rb)^2, 
\end{equation}
and:
\begin{equation}
	|\mg\force\md|=\frac{\mg q'^2\md }{4\pi\dens \son ^3} U \sim \frac{18 \pi \gamma^2 p_0^2 \alpha^2 r_b^2}{\rho_0 c_0^3} U. \label{eq:radi}
\end{equation}
For small bubbles, it is interesting to compare this radiation force to the Stokes drag:
\begin{equation}
	|\mathbf{F}_{\text{Sto}}|= C \pi \visc \rb \vit, \label{eq:stokes}
\end{equation}
with $C = 4$ for a bubble in a pure liquid moving at low Reynolds number, $C = 6$ if the surface is polluted so that the slip boundary condition is turned into a no slip boundary condition (see e.g. the book \cite{b_kim_2005}) and $C = 12$ for an undeformed bubble at large Reynolds numbers (see \cite{jfm_moore_1963}). If we compare equation \eqref{eq:radi} to \eqref{eq:stokes}, we obtain:
\begin{equation}
	\frac{|\mg\force\md|}{|\mathbf{F}_{\text{Sto}}|} \sim 
	=\frac{18}{C} \frac{\gamma^2 p_0^2\alpha^2 }{\visc\dens \son ^3} \rb.
\end{equation}
For a bubble rising in water at ambient temperature and pressure, we have $\dens \sim $\SI{1e3}{\kg\per\cubic\meter}, $\visc \sim $\SI{1e-3}{\pascal\s}, $\son \sim $\SI{1.5e3}{\meter\per\s}, $\gamma\sim\num{1.4}$, $p_0=$\SI{1e5}{\pascal}, so that with $\alpha\sim0.5$, we obtain:
\begin{equation}
	\frac{|\mg\force\md|}{|\mathbf{F}_{\text{Sto}}|} \sim 7 r_b.
\end{equation}
Hence, the self radiation force would be small compared to the Stokes drag for a millimetric bubble. However the self radiation force could become significant in cryogenic liquids such as liquid nitrogen or superfluid helium. In liquid nitrogen at $T\simeq\SI{77}{\kelvin}$, $\dens=$\SI{8e2}{\kg\per\cubic\meter}, $\visc=$\SI{2e-4}{\pascal\s}, $\son =$\SI{8e2}{\meter\per\s}, $\gamma\sim\num{1}$, $p_0=$\SI{1e5}{\pascal}, so that for $\alpha\sim0.5$ we obtain: \begin{equation}
	\frac{|\mg\force\md|}{|\mathbf{F}_{\text{Sto}}|} \sim 140 r_b,
\end{equation}
which means that for a bubble of a few millimeters in radius, the two phenomena would be of the same orders of magnitude. Note first that this calculation constitutes a rough comparison of the radiation force and Stokes drag since (i) for large bubble oscillations, the bubble dynamics becomes nonlinear, and (ii) the Stokes drag is also modified by the bubble oscillations as demonstrated by \cite{pof_magnaudet_1998}. Note also that the case of bubbles moving in a liquid constitutes just one possibility over the many configurations covered by equation \eqref{eq:final} since the above theory applies for an arbitrary monopolar source moving in an arbitrary fluid as soon as (i) the monopolar source emits a signal in the long wavelength regime and (ii) that the source is moving at low speed compared to the sound speed.

\section{Conclusion}

In this paper, we calculated the theoretical expression of the radiation force exerted on a translating monopolar source by its own acoustic field. We showed that the asymmetry of the radiated field due to the motion of the source creates a self-induced radiation force opposite to its motion. This first calculation opens many perspectives to calculate this effect for different types of sources, beyond the long wavelength regime and potentially unveil situations wherein this force is strong and could be in the opposite direction leading to a propulsive force, and hence with a scatterer surfing on its own acoustic field.

\section*{Acknowledgements}
We acknowledge support from ISITE ERC Generator program and stimulating discussions with Pr. J. Bush, which motivated us to perform this work.

\section*{Declaration of Interests}

 The authors report no conflict of interest.


\end{document}